\documentclass[twocolumn,english,conference]{IEEEtran}
\usepackage[T1]{fontenc}
\usepackage[latin9]{inputenc}
\usepackage{color}
\usepackage{float}
\usepackage{bm}
\usepackage{amsmath}
\usepackage{amssymb}
\usepackage{graphicx}

\makeatletter

\providecommand{\tabularnewline}{\\}
\floatstyle{ruled}
\newfloat{algorithm}{tbp}{loa}
\providecommand{\algorithmname}{Algorithm}
\floatname{algorithm}{\protect\algorithmname}

\usepackage{bm}
\usepackage{subfigure}
\usepackage{cite}
\usepackage[english]{babel}
\usepackage[T1]{fontenc}
\usepackage{algorithm}
\usepackage{algorithmic}

\IEEEoverridecommandlockouts

\makeatother

\usepackage{babel}
\begin{document}
\title{Deep Reinforcement Learning for Fresh Data Collection in UAV-assisted
IoT Networks }
\author{\IEEEauthorblockN{Mengjie~Yi\IEEEauthorrefmark{1}\IEEEauthorrefmark{2}, Xijun~Wang\IEEEauthorrefmark{3}\IEEEauthorrefmark{4},
Juan Liu\IEEEauthorrefmark{5}, Yan~Zhang\IEEEauthorrefmark{1}\IEEEauthorrefmark{2},
and Bo Bai\IEEEauthorrefmark{6}}\IEEEauthorblockA{\IEEEauthorrefmark{1}State Key Lab of Integrated Service Networks,\\
Information Science Institute, Xidian University, Xi\textquoteright an,
Shaanxi, 710071, China\\
\IEEEauthorrefmark{2}Science and Technology on Communication Network
Laboratory, Shijiazhuang, Hebei, 050081, China\\
\IEEEauthorrefmark{3}School of Electronics and Communication Engineering,
Sun Yat-sen University, Guangzhou, 510006, China\\
\IEEEauthorrefmark{4}Key Laboratory of Wireless Sensor Network \&
Communication, \\
Shanghai Institute of Microsystem and Information Technology, \\
Chinese Academy of Sciences, 865 Changning Road, Shanghai 200050 China\\
\IEEEauthorrefmark{5}School of Electrical Engineering and Computer
Science, Ningbo University, Zhejiang 315211, China\\
\IEEEauthorrefmark{6}Theory Lab (FKA Future Network Theory Lab),
2012 Labs, Huawei Technologies Co., Ltd., Hong Kong\\
Email: mjyi@stu.xidian.edu.cn, wangxijun@mail.sysu.edu.cn, eeliujuan@gmail.com,
\\
yanzhang@xidian.edu.cn, ee.bobbai@gmail.com }\thanks{This work was supported in part by the National Natural Science Foundation
of China (61971249), by Fundamental Research Funds for the Central
Universities under grant 19lgpy79, by the Research Fund of the Key
Laboratory of Wireless Sensor Network \& Communication (Shanghai Institute
of Microsystem and Information Technology, Chinese Academy of Sciences)
under grant 20190912, and by Young Stars in Science and Technology
of Shaanxi Province (2019KJXX-030).}}
\maketitle
\begin{abstract}
Due to the flexibility and low operational cost, dispatching unmanned
aerial vehicles (UAVs) to collect information from distributed sensors
is expected to be a promising solution in Internet of Things (IoT),
especially for time-critical applications. How to maintain the information
freshness is a challenging issue. In this paper, we investigate the
fresh data collection problem in UAV-assisted IoT networks. Particularly,
the UAV flies towards the sensors to collect status update packets
within a given duration while maintaining a non-negative residual
energy. We formulate a Markov Decision Process (MDP) to find the optimal
flight trajectory of the UAV and transmission scheduling of the sensors
that minimizes the weighted sum of the age of information (AoI). A
UAV-assisted data collection algorithm based on deep reinforcement
learning (DRL) is further proposed to overcome the curse of dimensionality.
Extensive simulation results demonstrate that the proposed DRL-based
algorithm can significantly reduce the weighted sum of the AoI compared
to other baseline algorithms.
\end{abstract}

\section{Introduction}

Owing to the fully controllable mobility and low operational cost,
unmanned aerial vehicles (UAVs) emerge as promising technologies to
provide wireless services \cite{mozaffariTutorialUAVsWireless2018}.
One of the most important applications is to collect information from
distributed sensors with the help of UAV in the Internet of Things
(IoT). Since the UAV can fly close to each sensor and exploit the
line-of-sight (LoS) dominant air-to-ground channel, the transmission
energy of the sensors can be greatly reduced and the throughput of
sensors can be significantly improved. Such advantages make UAV-assisted
IoT networks attract extensive attention in recent years and arouse
many research interests, ranging from the designs of UAV's flight
trajectory to resource allocation, and sensors' wakeup schedule \cite{zengEnergyEfficientUAVCommunication2016,zengEnergyMinimizationWireless2018,liuEnergyEfficientUAVControl2018,gongFlightTimeMinimization2018,challitaDeepReinforcementLearning2018}.
However, most of the existing works aimed at either maximizing system
throughput or minimizing delay. Recently, the age of information (AoI)
has been introduced to measure data freshness in IoT networks \cite{kaulRealtimeStatusHow2012,sunUpdateWaitHow2017,jiangTimelyStatusUpdate2018}.
Particularly, AoI tracks the time elapsed since the latest received
packet at the destination was generated at the source. In contrast
to throughput and delay, the AoI metric is defined from the receiver\textquoteright s
perspective. Therefore, previous results in the literature can not
be directly used to minimize the AoI in UAV-assisted IoT networks.

There have been some recent efforts on guaranteeing data freshness
in UAV-aided data collection for IoT networks. In \cite{abd-elmagidAveragePeakAgeofInformation2019},
the UAV was used as a mobile relay for a source-destination pair and
the trajectory is designed to minimize the average Peak AoI. In an
IoT network with multiple sensors, two age-optimal trajectory planning
algorithms were proposed in \cite{liuAgeoptimalTrajectoryPlanning2018},
where the UAV flies to and hovers above each sensor to collect data.
This work was then extended in \cite{tongUAVEnabledAgeOptimalData2019},
where the UAV collects data from a set of sensors when hovering at
each collection point (CP). The sensor-CP association and the UAV's
flight trajectory were jointly designed to minimize the maximum AoI
of the sensors. In a similar setup, an AoI deadline was imposed on
each sensor and the UAV's flight trajectory was designed to minimize
the number of expired packets in \cite{liMinimizingPacketExpiration2019}.
In these works, however, the UAV collects the data of each sensor
only once and then flies back to the depot. To continuously collect
data packets during a period of time, the authors of \cite{abd-elmagidDeepReinforcementLearning2019}
optimized both the UAV's flight trajectory and the transmission scheduling
of sensors to achieve the minimum weighted sum of AoI. Nonetheless,
the energy consumption of the UAV has not been considered in the design
of UAV's age-optimal trajectory. 

In this paper, by taking the energy constraint of UAV into consideration,
we study the age-optimal data collection problem in UAV-assisted IoT
networks based on deep reinforcement learning (DRL). In particular,
a UAV is dispatched from a depot, flies towards the sensors to collect
status update packets, and arrives at the destination within a given
duration. The UAV has to maintain a non-negative residual energy while
minimizing the weighted sum of the AoI of sensors during the flight.
To find the optimal flight trajectory of the UAV and transmission
scheduling of the sensors, we formulate this problem into a finite-horizon
Markov decision process (MDP). Due to the high-dimensional state space,
it is computationally prohibitive to solve the MDP problem using dynamic
programming algorithms. To address this issue, we propose a DRL-based
UAV-assisted data collection algorithm, where the UAV decides which
direction to fly and which sensor to connect at each step. Extensive
simulation results demonstrate that the proposed algorithm can significantly
reduce the weighted sum of AoI compared to other baseline policies.

The rest of this paper is organized as follows: The system model and
problem formulation are described in Section \ref{sec:System-Model}.
Section \ref{sec:Algorithm} provides the MDP formulation of the problem
and presents the proposed DRL-based algorithm. The simulation results
and discussions are given in Section \ref{sec:Simulation-Results}.
Finally, we conclude this paper in Section \ref{sec:Conclusions}.

\section{System Model and Problem Formulation\label{sec:System-Model}}

\subsection{Network Description}

\begin{figure}
\centering

\includegraphics[width=0.5\textwidth]{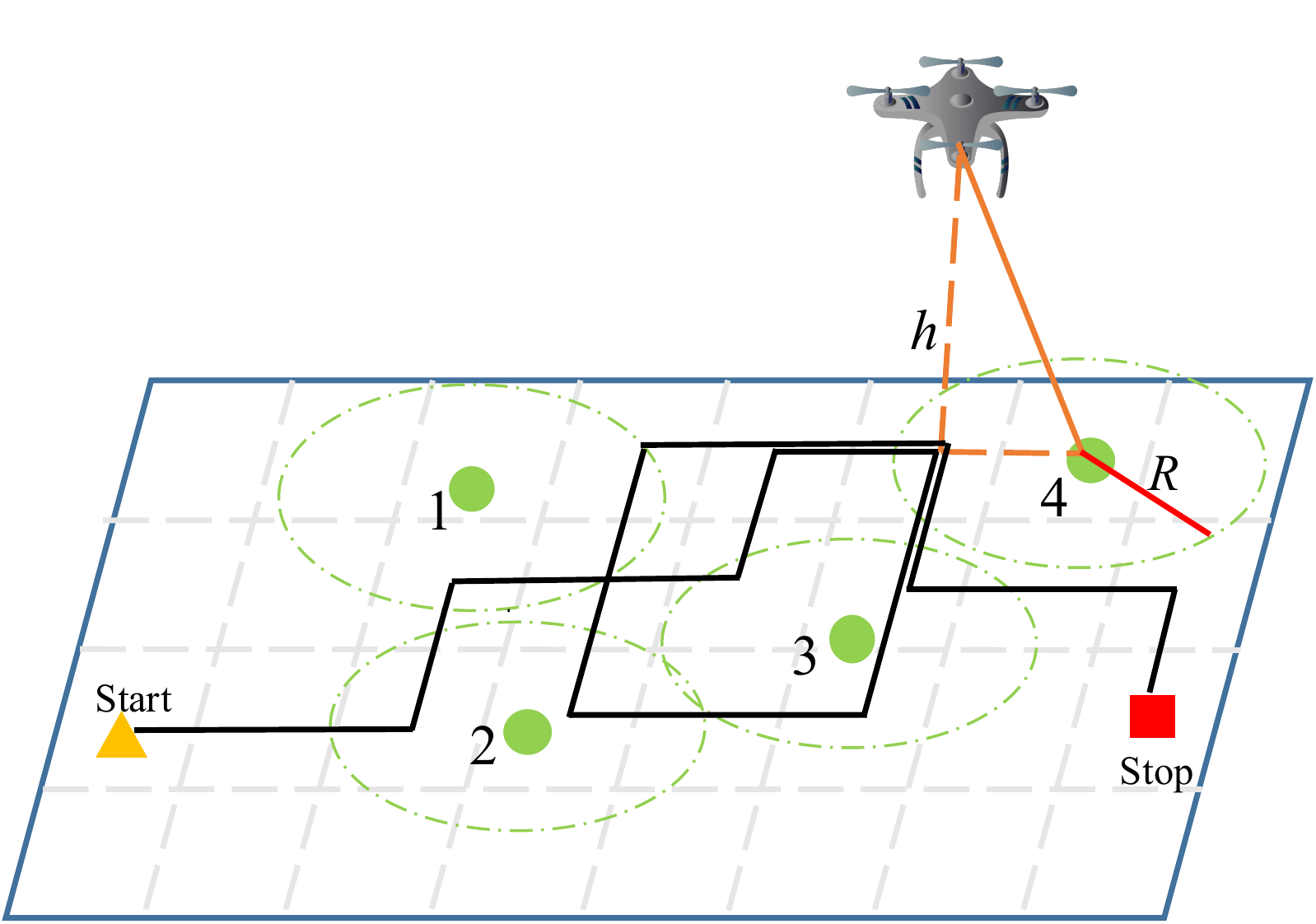}\caption{\label{fig:system-model}An illustration of the UAV-assisted data
collection.}
\end{figure}

As shown in Fig. \ref{fig:system-model}, we consider a UAV-assisted
IoT network, where $N$ sensor nodes (SNs) are randomly distributed
in a certain geographical region. The set of all the SNs is denoted
by $\mathcal{N}=\{1,2,\ldots,N\}$ and the location of each SN is
represented by $w_{n}=(x_{n},y_{n})$ for $n\in\mathcal{N}$. The
region of interest is equally partitioned into a number of small-size
grids such that the UAV\textquoteright s location is approximately
constant within each grid. Moreover, the center of the $l$-th grid
is represented by $c_{l}=(x_{l},y_{l})$. We denote by $\mathcal{C}$
the set containing the locations of centers for all the grids. Moreover,
the spacing distance between the centers of any two adjacent grids
is denoted by $L'$.

We assume a discrete-time system where time is divided into equal-length
time slots. The length of each slot is $\tau$ seconds. Given a time
duration of $T$ slots, the rotary-wing UAV takes off from an initial
location $c_{\text{start}}$ and flies over $N$ SNs to collect data
packets. At the end of the $T$-th slot, the UAV lands on a final
destination $c_{\text{stop}}$. We assume that the UAV flies along
the center of the grids at a fixed altitude $h$. In each time slot,
the UAV could hover over a certain grid or fly across one grid at
a constant speed $V$. Let $o_{t}$ denote the projection of the
UAV's location on the ground at time slot $t$. Then, the projection
of the UAV's flight trajectory is defined as a sequence of center
of grids $\bm{p}=(o_{1},o_{2},\ldots,o_{T})$, where $o_{1}=c_{\text{start}}$
and $o_{T}=c_{\text{stop}}$.

Let $E_{\text{max}}$ denote the initial amount of energy the UAV
carries. The energy consumption of the UAV consists of the communication
energy and the propulsion energy. Since the communication energy consumption
is relatively small, we consider only the propulsion energy consumption
in this paper. The propulsion energy of the rotary-wing UAV is mainly
composed of the blade profile energy, the induced power, and the parasite
energy \cite{zengEnergyMinimizationWireless2018}. The propulsion
power consumption can be expressed as follows,
\begin{align}
\tilde{P}(V_{t})= & P_{0}\left(1+\frac{3V_{t}^{2}}{U_{tip}^{2}}\right)+P_{1}\left(\sqrt{1+\frac{V_{t}^{4}}{4v_{0}^{4}}}-\frac{V_{t}^{2}}{2v_{0}^{2}}\right)^{\frac{1}{2}}\nonumber \\
 & +\frac{1}{2}d_{0}\rho s_{0}AV_{t}^{3},
\end{align}
where $P_{0}$ and $P_{1}$ represent the blade profile power and
derived power of the UAV in the hovering state, respectively, $V_{t}$
is the velocity of the UAV at slot $t$, $U_{tip}$ represents the
tip speed of the rotor blade of the UAV, $v_{0}$ represents the mean
rotor induced velocity in the hovering state, $d_{0}$ is the fuselage
drag ratio, $\rho$ represents the density of air, $s_{0}$ indicates
the rotor solidity, and $A$ represents the area of the rotor disk.
In particular, the power consumption when hovering (i.e., $V_{t}=0$)
is $\tilde{P}(0)=P_{0}+P_{1}$.

We assume that the UAV could establish the LoS links with the SNs
due to its high attitude. Then, the channel power gain from the SN
to the UAV at time slot $t$ can be given by 
\begin{equation}
g_{n,u}(t)=\beta_{0}d_{n,u}^{-2}(t)=\frac{\beta_{0}}{\parallel o_{t}-w_{n}\parallel^{2}+h^{2}},
\end{equation}
where $\beta_{0}$ is the channel gain at a reference distance of
$1$ meter, $d_{n,u}(t)$ denotes the Euclidean distance between the
SN $n$ and the UAV at time slot $t$. Let $P$ denote the transmission
power of each SN. When the UAV is within the coverage of one SN, i.e.,
$\parallel o_{t}-w_{n}\parallel\leq R$, the SN generates a status
update of size $M$  and sends it to the UAV successfully in a time
slot. Specifically, the coverage radius can be calculated as
\begin{equation}
R=\left(\frac{\beta_{0}P}{(2^{\frac{M}{B\tau}}-1)\sigma^{2}}-h^{2}\right)^{\frac{1}{2}},
\end{equation}
where $B$ is the channel bandwidth, and $\sigma^{2}$ is the noise
power at the UAV.

We employ AoI to measure the freshness of information. In particular,
the AoI is defined as the time elapsed since the generation of the
latest status update received by the UAV. Let $U_{n}(t)$ denote the
time at which the latest status update of SN $n$ successfully received
by the UAV was generated. The AoI of SN $n$ at the beginning of slot
$t$ is then given by
\begin{equation}
\delta_{n,t}=t-U_{n}(t).\label{eq:AoI}
\end{equation}
Let $\bm{b}=(b_{1},b_{2},\ldots,b_{T})$ be the vector of the SNs'
scheduling variables, where $b_{t}\in\mathcal{B}\triangleq\{0,1,\ldots,N\}$
denotes which SN is scheduled to update its status at time slot $t$.
In particular, $b_{t}=n$ indicates that SN $n$ transmits to the
UAV at slot $t$ and $b_{t}=0$ means that no transmission occurs
at slot $t$. According to (\ref{eq:AoI}), if SN $n$ is scheduled
to transmit at slot $t$ and the UAV is located in the coverage of
SN $n$, then its AoI decreases to one; otherwise, the AoI increases
by one. Then, the dynamics of the AoI can be given by
\begin{equation}
\delta_{n,t+1}=\begin{cases}
1, & \text{if }b_{t}=n\text{ and }\parallel o_{t}-w_{n}\parallel\leq R;\\
\delta_{n,t}+1, & \text{otherwise}.
\end{cases}\label{eq:Dynamic-AoI}
\end{equation}

\subsection{Problem Formulation}

Our objective is to find the optimal trajectory of the UAV and the
optimal scheduling of the SNs that minimize the weighted average AoI
of all the SNs. The optimization problem can be expressed as follows: 

\begin{align}
\text{P1: }\underset{\bm{p},\bm{b}}{\min} & \frac{1}{T}\sum_{t=1}^{T}\sum_{n=1}^{N}\theta_{n}\delta_{n,t},\label{eq:object function}\\
s.t. & \sum_{t=1}^{T}\tilde{P}(V_{t})\tau\leq E_{\text{max}},\label{eq:energy}\\
 & o_{1}=c_{\text{start}},\label{eq:start pos}\\
 & o_{T}=c_{\text{stop}}.\label{eq:stop pos}
\end{align}
where $\theta_{n}$ denotes the importance of SN $n$. (\ref{eq:energy})
ensures that the UAV will not run out of the energy before time slot
$T$. (\ref{eq:start pos}) and (\ref{eq:stop pos}) guarantee that
the UAV starts from the initial location and arrives at the final
location at time slot $T$. It is easily observed that the above optimization
problem is a nonlinear integer programming one, which is computationally
complex to solve for large-scale networks. In the following section,
we propose a learning based algorithm for the UAV to learn its trajectory
and the SNs' transmission schedule at each location along the trajectory. 

\section{DRL-based Approach \label{sec:Algorithm}}

In this section, we first cast the UAV-assisted data collection problem
into a Markov decision process (MDP) and then propose a DRL-based
algorithm to minimize the weighted average AoI of all the SNs.

\subsection{MDP Formulation}

We reformulate the problem P1 via an MDP, which is usually represented
by a tuple $(s,a,r,p)$. Here, $s$ presents the state, $a$ denotes
the action, $r$ is the reward function, and $p$ is the state transition
probability. MDP is commonly used to model a sequential decision-making
process. In particular, at time slot $t$, the agent observes some
state $s_{t}$ and performs an action $a_{t}$. After taking this
action, the state of the environment transits to $s_{t+1}$ with probability
$p_{s_{t},s_{t+1}}$, and the agent receives a reward $r_{t}$. We
consider the UAV as the agent for performing the data collection algorithm
and define the state, action, and reward function in the following.

\subsubsection{State}

The state at time slot $t$ is defined as $s_{t}=(o_{t},\bm{\delta}_{t},\phi_{t},\Delta_{t})$,
which is composed of four parts:
\begin{itemize}
\item $o_{t}\in\mathcal{C}$ is the projection of the UAV on the ground
at time slot $t$.
\item $\bm{\delta}_{t}=(\delta_{1,t},\delta_{2,t},\ldots,\delta_{N,t})$
is the AoI of all the SNs at the UAV at time slot $t$. For the AoI
of each SN, we have $\delta_{n,t}\in\mathcal{D}\triangleq\{1,2,\ldots,\delta_{\text{max}}\}$,
where $\delta_{\text{max}}$ is the maximum value of AoI and can be
chosen to be arbitrary large.
\item $\phi_{t}\in\mathcal{T}\triangleq\{0,1,\ldots,T\}$ is the difference
between the remaining time of the UAV and the minimum time required
to reach the final destination.
\item $\Delta_{t}=E_{\text{max}}-\sum_{m=1}^{t}\tilde{P}(V_{t})\tau-\sum_{m=1}^{N-t}\tilde{P}(V)\tau$
is the difference between the remaining energy of the UAV and the
energy required for the UAV to arrive at the final destination in
the remaining time. $\Delta_{t}\in\mathcal{E}$, where $\mathcal{E}$
is the set of the energy level of the UAV.
\end{itemize}
Altogether, the state space of the system can be expressed as $\mathcal{S}=\mathcal{C}\times\mathcal{D}^{N}\times\mathcal{T}\times\mathcal{E}$.

\subsubsection{Action}

The action of the UAV at time slot $t$ is characterized by its movement
$\upsilon_{t}$ and the scheduling of SN $b_{t}$, i.e., $a_{t}=(\upsilon_{t},b_{t})$.
In each time slot, the UAV either hovers at its current location or
move to one of its adjacent cells. Specifically, $\upsilon_{t}\in\mathcal{V}\triangleq\{\text{North},\text{South},\text{East},\text{West},\text{Hovering}\}$.
Then, the action space is given by $\mathcal{A}=\mathcal{V}\times\mathcal{B}$.

\subsubsection{Reward}

In our context of the UAV-assisted data collection, the reward should
encourage the UAV to minimize the weighted average AoI of all the
SNs under the constraints given by (\ref{eq:energy})-(\ref{eq:stop pos}).
When the UAV reaches the final destination at time slot $T$ with
a non-negative residual energy, we will give the UAV an additional
reward. However, a punishment will be imposed when the constraints
are violated. Let $J=\frac{1}{T}\sum_{t=1}^{T}\sum_{n=1}^{N}\theta_{n}\delta_{n,t}$.
Then, the reward is defined as follows,\textcolor{black}{
\begin{equation}
r_{t}=\begin{cases}
-J-k_{1}, & \textrm{if }\phi_{t}<0,\\
-J-k_{2}, & \textrm{if }\Delta_{t}<0,\\
-J+k_{3}, & \textrm{if }o_{T}=c_{\text{stop}},\Delta_{t}\geq0,\\
-J, & \textrm{otherwise.}
\end{cases}\label{eq:reward}
\end{equation}
where $k_{1}$, $k_{2}$, and $k_{3}$ are positive constants and
set large enough. }

\subsubsection{State Transition}

The AoI of each SN is updated as in (\ref{eq:Dynamic-AoI}). The dynamics
of the UAV\textquoteright s location can be expressed as 
\begin{equation}
o_{t+1}=\begin{cases}
o_{t}+(0,L'), & \text{if }\upsilon_{t}=\text{North},\\
o_{t}-(0,L'), & \text{if }\upsilon_{t}=\text{South},\\
o_{t}+(L',0), & \text{if }\upsilon_{t}=\text{East},\\
o_{t}-(L',0), & \text{if }\upsilon_{t}=\text{West},\\
o_{t}, & \text{if }\upsilon_{t}=\text{Hovering}.
\end{cases}
\end{equation}

The time difference $\phi_{t}$ is updated based on the UAV's location.
In particular, if the UAV flies towards the final destination at slot
$t$, $\phi_{t+1}$ remains the same as $\phi_{t}$. If the UAV hovers
at slot $t$, $\phi_{t+1}$ is decreased by one. While $\phi_{t+1}$
is decreased by two, if the UAV flies away from the final destination.
Altogether, we can update $\phi_{t}$ as follows,
\begin{equation}
\phi_{t+1}=\begin{cases}
\phi_{t}, & \text{if }\parallel o_{t}-c_{\text{stop}}\parallel>\parallel o_{t+1}-c_{\text{stop}}\parallel,\\
\phi_{t}-1, & \text{if }\parallel o_{t}-c_{\text{stop}}\parallel=\parallel o_{t+1}-c_{\text{stop}}\parallel,\\
\phi_{t}-2, & \text{if }\parallel o_{t}-c_{\text{stop}}\parallel<\parallel o_{t+1}-c_{\text{stop}}\parallel.
\end{cases}
\end{equation}

Since the power consumptions for hovering and flying are different,
the update of energy difference $\Delta_{t}$ is different for these
two cases. According to the definition of the energy difference $\Delta_{t}$,
the update of $\Delta_{t}$ can be given by
\begin{equation}
\Delta_{t+1}=\begin{cases}
\Delta_{t}+\tilde{P}(V)-\tilde{P}(0), & \text{\text{if} }\upsilon_{t}=\text{Hovering},\\
\Delta_{t} & \text{otherwise}.
\end{cases}
\end{equation}

Our goal is to find an age-optimal policy $\pi^{*}$, which determines
the sequential actions over a finite horizon of length $T$. Given
a policy $\pi$, the total expected reward of the system starting
from an initial state $s_{1}$ is defined as
\begin{equation}
G_{\pi}=\sum_{t=1}^{T}\mathbb{E}_{\pi}\left[r_{t}\mid s_{1}\right].
\end{equation}
Then, the optimal policy can be obtained by maximizing the total expected
reward, i.e., $\pi^{*}=\arg\max\limits _{\pi}G_{\pi}$. When the number
of SNs become large, it is computationally infeasible to find the
optimal strategy by standard dynamic programming method. Therefore,
DRL is employed in the following subsection to solve this problem. 

\subsection{DRL Approach}

We employ DQN, which is one of the most well adopted DRL method, to
derive the optimal policy. In this approach, we define a state-action
value function $Q_{\pi}(s,a)$, which represents the expected reward
for selecting action $a$ in state $s$ and then following policy
$\pi$. The optimal Q-value function can be estimated by the update
\begin{align}
Q(s_{t},a_{t})= & Q(s_{t},a_{t})+\nonumber \\
 & \alpha\left[r_{t}+\max_{a}Q(s_{t+1},a)-Q(s_{t},a_{t})\right],
\end{align}
where $\alpha$ is the learning rate. The optimal policy is the one
that takes the action which maximizes the Q-value function at each
step.

 By incorporating deep neural network (DNN) into the framework of
Q-learning, DQN can overcome the curse of dimensionality. In particular,
we use a DNN with weights $\theta$ to approximate the Q-value function
$Q(s,a)$ with $Q(s,a;\theta)$. The DNN can be trained by minimizing
a sequence of loss function $L(\theta_{t})$ that changes at each
slot $t$. Specifically, 
\begin{equation}
L(\theta_{t})=\left(r_{t}+\max_{a}Q(s_{t+1},a;\theta_{t-1})-Q(s_{t},a_{t};\theta_{t})\right)^{2},
\end{equation}
where the weights are updated at slot $t$ and the weight $\theta_{t-1}$
from the previous slot are held fixed. However, the use of one DNN
may induce instability. In order to overcome this issue, two neural
networks are employed \cite{mnihHumanlevelControlDeep2015}, i.e.,
the current network with weights $\theta$ and the target network
parameterized by $\theta^{-}$. The current network is used as a function
approximator and its weights are updated at every slot. While the
target network computes the target Q-value function and its weights
are fixed for a while and updated at every $O$ steps (Lines 11\textasciitilde 18).
In particular, the weights of the DNN are updated by minimizing the
loss function, which is defined as 
\begin{align}
L(\theta)= & \left(r_{m}+\max\limits _{a}Q(s_{m+1},a;\theta^{-})-Q(s_{m},a_{m};\theta)\right)^{2},\label{eq:Loss-Func}
\end{align}
where $Q(s_{m},a_{m};\theta)$ is evaluated by the current network
and $Q(s_{m+1},a;\theta^{-})$ is evaluated by the target network.
Based on this, the update formula for weights $\theta$ is given as
follows:
\begin{align}
\theta= & \theta+\alpha[y_{m}-Q(s_{m},a_{m};\theta)]\nabla_{\theta}Q(s_{m},a_{m};\theta),\label{eq:Gradient}
\end{align}
where $y_{m}=r_{m}+\max\limits _{a}Q(s_{m+1},a;\theta^{-})$ and $\nabla_{\theta}$
denotes the gradient with respect to $\theta$. 

\begin{algorithm}[tb]
\caption{DRL-based UAV-assisted data collection algorithm \label{alg:DP_AoI_optimal}}
\begin{algorithmic}[1]

\STATE Initialize the replay memory $D$, the probability $\epsilon$,
the current network parameter $\theta$, and the target network parameter
$\theta^{-}=\theta$;

\STATE Initialize the current network $Q(s,a;\theta)$ with weights
$\theta$ and the target network $Q(s,a;\theta^{-})$ with weights
$\theta^{-}$;

\FOR{ $episode=1:E$ }

\STATE Initialize the environment and observe an initial state $s_{1}$;

\FOR{$t=1:T$}

\STATE Select a random action $a_{t}$ with probability $\epsilon$;

\STATE Otherwise select $a_{t}=\arg\max\limits _{a}Q(s_{t},a;\theta)$;

\STATE Execute action $a_{t}$ and observe the reward $r_{t}$ and
the next state $s_{t+1}$;

\STATE Mark $s_{n+1}$ if it is a terminal state and store transition
$(s_{t},a_{t},r_{t},s_{t+1})$ in the replay memory;

\STATE Sample a random mini-batch of transitions $(s_{m},a_{m},r_{m},s_{m+1})$
from the replay memory;

\STATE Calculate the target value $y_{m}$:

\IF{ $s_{m+1}$ is the terminal state}

\STATE $y_{m}=r_{m}$;

\ELSE

\STATE $y_{m}=r_{m}+\max\limits _{a}Q(s_{m+1},a;\theta^{-})$;

\ENDIF

\STATE Update the current network by performing the gradient descent
in (\ref{eq:Gradient});

\STATE Update target parameters, $\theta^{-}=\theta$, in every $O$
steps;

\STATE Terminate the episode if $s_{m+1}$ is the terminal state.

\ENDFOR

\ENDFOR

\end{algorithmic}
\end{algorithm}

Based on the DQN with two neural networks, the \textcolor{black}{UAV-assisted
data collection} algorithm is proposed to find the optimal solution
to problem P1, and the details are showed in Algorithm \ref{alg:DP_AoI_optimal}.
At the beginning of the training process, the estimation of the Q-value
function is far from accurate. Hence, the UAV should explore the environment
more often at first. When the policy continues improving and the knowledge
of the environment is more accurate, the UAV should exploit the learned
knowledge more often. As such, we utilize a simple $\epsilon$-greedy
policy (Lines 6\textasciitilde 7). In particular, the action is randomly
selected to explore the environment with probability $\epsilon$ and
the action that maximizes $Q(s_{t},a;\theta)$ is chosen to exploit
the policy with probability $1-\epsilon$. Moreover, $\epsilon$ is
set to be decreasing with the number of slots so that the UAV can
choose the optimal action when the estimation of Q-value function
converges. 

Experience replay is used in the learning process. The agent stores
the experience $(s_{t},a_{t},r_{t},s_{t+1})$ in the replay memory,
and then samples a mini-batch of the experiences from the replay memory
uniformly at random to train the neural network (Lines 9\textasciitilde 10).
By using experience replay, not only the correlation among the continuous
samples is reduced, but also the utilization rate of the experience
data can be improved. \textcolor{black}{We also note that the UAV-assisted
data collection problem we considered is episodic, since the UAV is
required to be arrive in the final destination at time slot $T$.
In particular, there are three terminal cases: 1) when the UAV reaches
the final destination at time slot $T$, 2) when $\phi_{t}<0$, and
3) when $\Delta_{t}<0$ (Line 19).}

\section{Simulation Results\label{sec:Simulation-Results}}

In this section, we perform extensive simulations to evaluate the
performance of the DRL-based UAV-assisted data collection algorithm
in an IoT network. We consider a square area of $500\text{ m}\times500\text{ m}$
that is virtually divided into $20\times20$ equally-sized grids of
length $25$ m. Let the center of the left lower grid of the square
region be the origin with coordinate $[0,0]$ and the index of every
grid is the coordinate of the grid center divided by 25. For instance,
the left lower grid is indexed by $(0,0)$. We assume that UAV\textquoteright s
initial and final locations are at grids $(10,0)$ and $(10,19)$,
respectively. We also assume that the SNs have equal importance weights.
Unless otherwise specified, the simulation parameters are presented
in Table I. 

The two neural networks in the proposed algorithm is implemented using
Tensorflow. In particular, each DNN includes two fully-connected hidden
layers with 200 and 256 neurons. The input layer size of the DNN is
the same as the state space size and the output layer size of the
DNN is equal to the total number of actions. The hypeparameters of
DQN are summarized in Table \ref{tab:Hyperparameters-of-DQN}.

\begin{table}
\caption{System parameters\label{tab:system-parameters}}

\centering

\begin{tabular}{c|c}
\hline 
Parameter & Value\tabularnewline
\hline 
Channel bandwidth $B$ & 1 MHz\tabularnewline
\hline 
Update size $M$ & 5 Mbits\tabularnewline
\hline 
Noise power $\sigma^{2}$ & -100 dbm\tabularnewline
\hline 
Channel gain at $1$ m $\beta_{0}$ & -60 dB\tabularnewline
\hline 
Flight altitude $h$ & $120$ m\tabularnewline
\hline 
Time duration $T$ & 70 slots\tabularnewline
\hline 
UAV speed $V$ & 25 m/s\tabularnewline
\hline 
Initial energy $E_{\text{max}}$ & 2.2e4 J\tabularnewline
\hline 
Air density in $\rho$ & 1.225 kg/m$^{3}$\tabularnewline
\hline 
Tip speed $U_{tip}$  & 120 m/s\tabularnewline
\hline 
Blade profile power $P_{0}$  & 99.66 W\tabularnewline
\hline 
Derived power $P_{1}$  & 120.16 W\tabularnewline
\hline 
Body resistance ratio $d_{0}$ & 0.48\tabularnewline
\hline 
Robustness of the rotor $s$ & 0.0001\tabularnewline
\hline 
The area of the rotor disk $A$  & 0.5 s$^{2}$\tabularnewline
\hline 
Mean rotor induced velocity in hover $v_{0}$  & 0.002 m/s\tabularnewline
\hline 
\end{tabular}
\end{table}

\begin{table}
\caption{Hyperparameters of DQN\label{tab:Hyperparameters-of-DQN}}

\centering

\begin{tabular}{c|c}
\hline 
Parameter & Value\tabularnewline
\hline 
Episodes $E$ & 20000\tabularnewline
\hline 
Reply memory size $D$ & 40000\tabularnewline
\hline 
Mini-batch size & 200\tabularnewline
\hline 
Initial $\epsilon$ & 0.9\tabularnewline
\hline 
$\epsilon$-greedy decrement & 0.0001\tabularnewline
\hline 
Minimum $\epsilon$ & 0\tabularnewline
\hline 
Learning rate $\delta$ & 0.002\tabularnewline
\hline 
Learning rate decay rate & 0.95\tabularnewline
\hline 
Learning rate decay step & 10000\tabularnewline
\hline 
Update step $O$ & 300\tabularnewline
\hline 
Optimizer & Adam\tabularnewline
\hline 
Activation function & ReLU\tabularnewline
\hline 
\end{tabular}
\end{table}

In the following figures, we compare the performance of the proposed
algorithm with two baseline algorithms, namely AoI-based algorithm
and distance-based algorithm. In the AoI-based algorithm, the UAV
flies to the SN with the largest AoI in the current time slot. While
in the distance-based algorithm, the flight trajectory of the UAV
is divided into multiple rounds. In each round, the UAV traverses
all the SNs one by one and the UAV flies to the nearest and unvisited
SN in the current traversal round. Moreover, the UAV can collect status
update from the SNs on its way in both baseline algorithms. When the
UAV\textquoteright s residual energy or the remaining time is less
than a threshold, it directly flies to the final destination.  

\begin{figure}[t]

\includegraphics[width=0.5\textwidth]{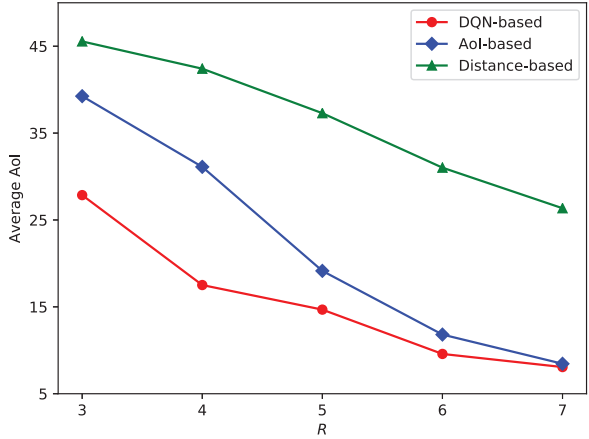}\caption{\label{fig:R_impact}Effect of $R$ on the average AoI with $N=3$.}
\end{figure}

Fig. \ref{fig:R_impact} illustrates the average AoI with respect
to the coverage radius $R$ in a scenario with three randomly deployed
SNs, where the value of $R$ is normalized by the length of a grid.
From Fig. \ref{fig:R_impact}, we can see that a higher $R$ results
in lower average AoI since it takes less time for the UAV to fly to
collect data packets. Moreover, we can see that our proposed DQN-based
algorithm outperforms the two baseline algorithms since it jointly
considers the AoI, the location of the UAV, and the time and energy
constraints. It is also shown that the AoI-based algorithm  achieves
almost the same performance as DQN-based algorithm when $R$ is large.
This is because there is an overlap of the coverage of all the SNs
for a larger $R$ and the UAV can fly above the overlapping area to
collect data packets. 

\begin{figure}
\centering

\includegraphics[width=0.5\textwidth]{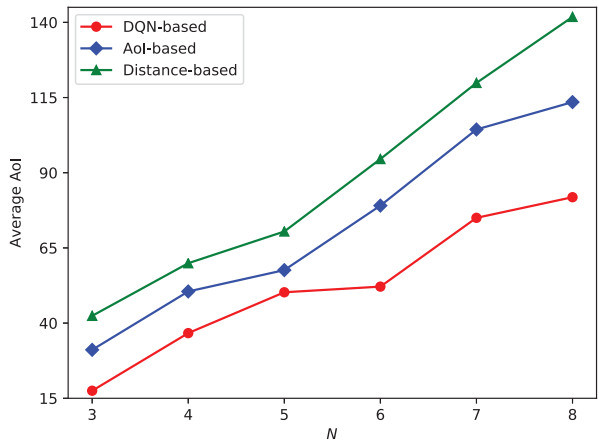}\caption{\label{fig:sensor_num} Effect of $N$ on the average AoI with $R=4$.}
\end{figure}

Fig.\ref{fig:sensor_num} shows the average AoI with respect to the
number of sensors $N$ for $R=4$. We can easily observe that by adopting
our DQN-based algorithm, the average AoI is smaller than that of the
baseline algorithms. Moreover, the reduction of the average AoI is
more significant for a larger $N$. Fig. \ref{fig:sensor_num} also
shows that the average AoI increases with the number of SNs. This
is because, for a larger $N$, the UAV has to fly farther to collect
update packets. In addition, the SNs have to wait for a longer time
to update their status, since the UAV can collect data packets from
only one SN each time.

\section{Conclusions\label{sec:Conclusions}}

In this paper, we have investigated the AoI-optimal data collection
problem in UAV-assisted IoT networks, where a UAV collects status
update packets and arrives at the final destination under both time
and energy constraints. In order to minimize the wighted sum of the
AoI, we have formulated the problem as a finite-horizon MDP. We have
then designed a DRL-based data collection algorithm to find the optimal
flight trajectory of the UAV and the transmission scheduling of the
SNs. Moreover, we have conducted extensive simulations and shown that
the DRL-based algorithm is superior to two baseline approaches, i.e.,
the AoI-based and the distance-based algorithms. Simulation results
also demonstrated that the weighted sum of the AoI is monotonically
decreasing with the SN's coverage radius and monotonically increasing
with the number of SNs. 

\appendices{}

\bibliographystyle{IEEEtran}
\bibliography{AoI}

\end{document}